\documentclass[sigconf]{acmart}
\usepackage{balance}
\usepackage{latexsym}
\usepackage{multicol}
\usepackage{multirow}
\usepackage{booktabs}
\usepackage{algorithm}
\usepackage{subfigure}
\usepackage{makecell}
\usepackage{enumitem}
\usepackage{algpseudocode}
\AtBeginDocument{%
  \providecommand\BibTeX{{%
    \normalfont B\kern-0.5em{\scshape i\kern-0.25em b}\kern-0.8em\TeX}}}

\copyrightyear{2026}
\acmYear{2026}
\setcopyright{cc}
\setcctype{by}
\acmConference[RecSys '26]{20th ACM Conference on Recommender Systems}{September 27-October 02, 2026}{Minneapolis, MN, USA}
\acmBooktitle{20th ACM Conference on Recommender Systems (RecSys '26), September 27-October 02, 2026, Minneapolis, MN, USA}
\acmDOI{10.1145/3773078.3831873}
\acmISBN{979-8-4007-2284-4/2026/09}

\begin{document}



\title{ConAlign: Conditional Alignment Framework for Balancing Biased and Unbiased Recommendation}

\renewcommand{\thefootnote}{\fnsymbol{footnote}}
\author{Jingcheng Zhang\footnotemark[1]\footnotemark[2]}
\email{zhangjingcheng@kuaishou.com}
\affiliation{%
  \institution{Kuaishou Technology}
  \city{Beijing}
  \country{China}
}

\author{Yihan Wang\footnotemark[1]}
\email{wangyihan05@kuaishou.com}
\affiliation{%
  \institution{Kuaishou Technology}
  \city{Beijing}
  \country{China}
}

\author{Qi Song}
\email{songqi@kuaishou.com}
\affiliation{%
  \institution{Kuaishou Technology}
  \city{Beijing}
  \country{China}
}

\author{Liyin Hong\footnotemark[3]}
\email{hongliyin@kuaishou.com}
\affiliation{%
  \institution{Kuaishou Technology}
  \city{Beijing}
  \country{China}
}

\renewcommand{\shortauthors}{Jingcheng Zhang et al.}

\begin{abstract}
Industry recommender systems trained on observational data suffer from various biases that create filter bubbles, causing user interests to collapse into narrow categories and severely degrading long-term engagement. While utilizing unbiased uniform data for debiasing has shown promise, existing methods remain impractical for industrial deployment due to limitations such as neglect of factual (biased) recommendation performance and the substantial computational overhead. To overcome these limitations, we propose ConAlign (Conditional Alignment Framework), a conditional debiasing approach for industrial deployment. The key innovation of ConAlign lies in a discrete gating-based conditional alignment mechanism that selectively transfers knowledge from the biased tower to the unbiased tower. Following a selective intervention paradigm rather than universal correction, it seamlessly balances factual accuracy and unbiased preference estimation while supporting real-time streaming adaptation. To the best of our knowledge, ConAlign is the first streaming debiasing recommendation framework successfully deployed in a large-scale industrial recommendation system that utilizes a small fraction of unbiased random traffic for debiasing. Extensive offline experiments on three real-world datasets rigorously validate the effectiveness of our proposed framework\footnotemark[4].\footnotetext[4]{Code is available at: \url{https://github.com/JcZhangzz/ConAlign}.} Furthermore, large-scale online A/B testing on Kuaishou demonstrates significant improvements in long-term user engagement and interest diversity, with negligible latency overhead.
\end{abstract}



\maketitle
\footnotetext[1]{These authors contributed equally to this work.}
\footnotetext[2]{Work done during an internship at Kuaishou Technology.}
\footnotetext[3]{Corresponding author.}
\renewcommand{\thefootnote}{\arabic{footnote}}
\setcounter{footnote}{0}

\thispagestyle{empty}

\section{Introduction}
\label{sec:Introduction}

\begin{figure*}[h]
    \centering
    \includegraphics[width=0.75\linewidth]{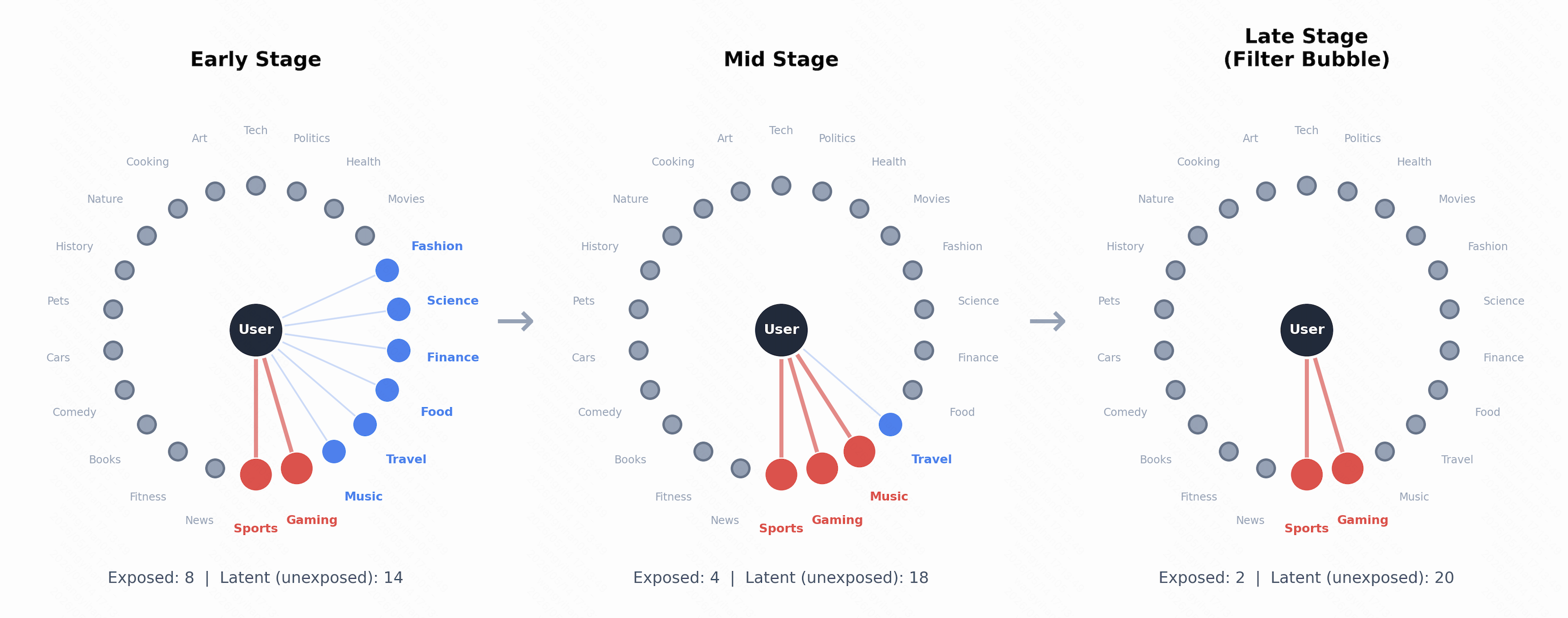}
    \caption{The filter bubble phenomenon in biased recommender systems. Nodes represent latent user interests; \textcolor{red}{red}: over-reinforced; \textcolor{blue}{blue}: occasionally exposed; grey: never surfaced. Over time, the system concentrates on a shrinking set of dominant interests, suppressing diversity and degrading long-term engagement.}
    \label{fig:filter_bubble}
\end{figure*}

\begin{figure}[h]
  \centering
  \includegraphics[width=0.75\linewidth]{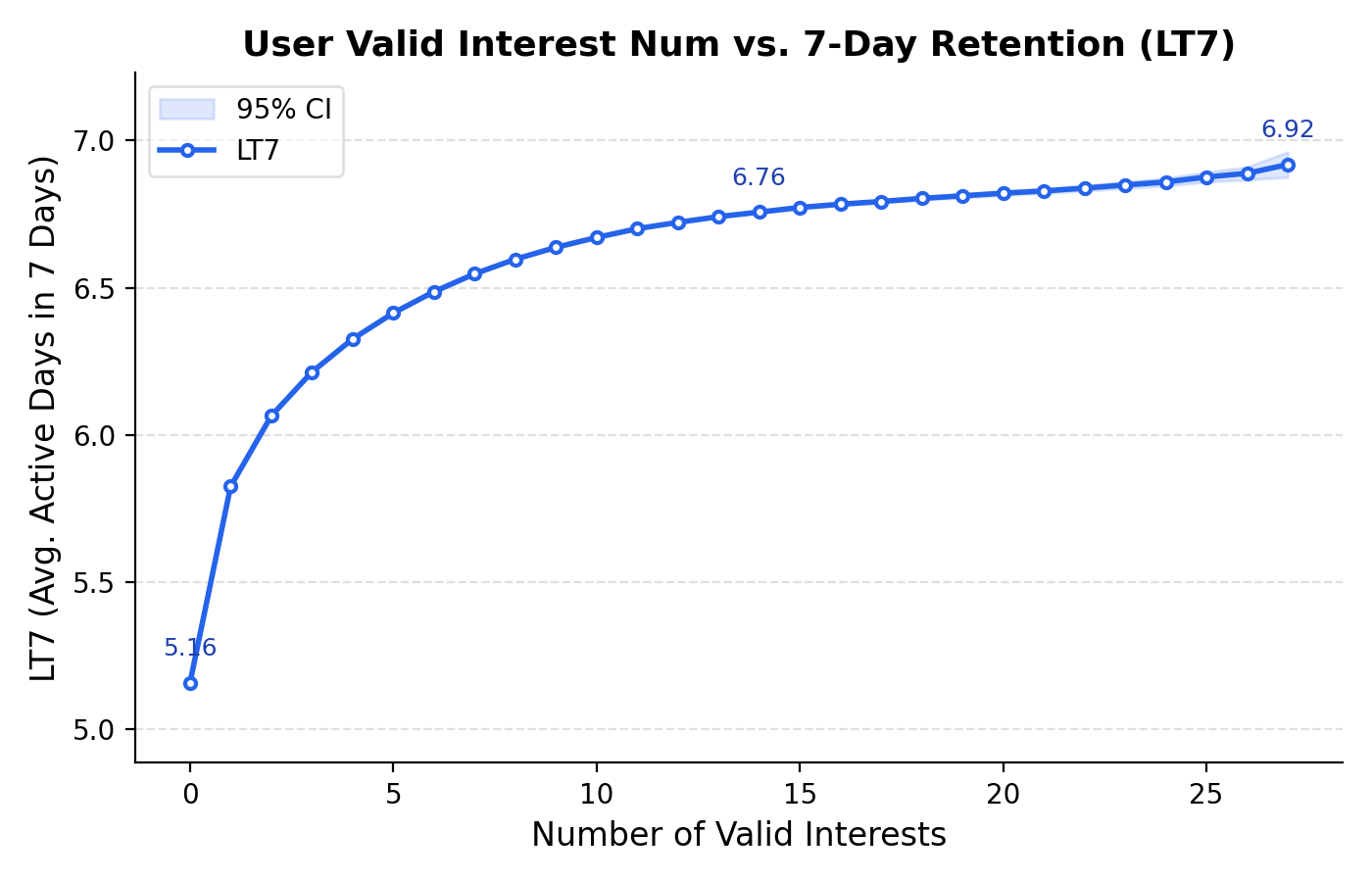}
  \caption{The correlation between Valid Interest Number (VIN), defined as the number of interest categories in which a user has accumulated sufficient positive feedback over 7 days, and the LT7 metric, based on online real-world data.}
  \label{fig:interests_LT7}
\end{figure}

Recommender systems have become the fundamental infrastructure for content discovery on modern online platforms. In practice, however, recommendation models are trained on logged interactions collected under historical serving policies rather than uniformly sampled user preferences. As a result, the observed feedback is inevitably entangled with various biases~\cite{chenBiasDebiasRecommender2023}, such as exposure bias~\cite{liangModelingUserExposure2016}, selection bias~\cite{marlinCollaborativeFilteringMissing2012}, popularity bias~\cite{krishnanAdversarialApproachImprove2018}, and position bias~\cite{collinsStudyPositionBias2018}. This mismatch causes recommendation models to overfit historically reinforced interaction patterns, which in turn leads to self-reinforcing feedback loops and filter bubbles~\cite{jiangDegenerateFeedbackLoops2019, mansouryFeedbackLoopBias2020, rowlandFilterBubbleWhat2011}, as illustrated in Figure~\ref{fig:filter_bubble}. Such filter bubbles not only reduce recommendation diversity, but also pose a long-term business risk for industrial recommender systems~\cite{zieglerImprovingRecommendationLists2005, kaminskasDiversitySerendipityNovelty2017}. Figure~\ref{fig:interests_LT7} further presents real-world evidence from Kuaishou, showing that user interest diversity is positively correlated with long-term retention (LT7). This observation highlights the importance of preserving diverse user interests for sustainable platform engagement.

To mitigate the impact of biased observational feedback, prior studies have explored causal correction~\cite{IPS,wangDoublyRobustJoint2019}, invariant learning~\cite{wangInvariantPreferenceLearning2022, KD-Debias}, and adversarial learning~\cite{zhuAdversarialPropensityWeighting} to recover bias-invariant user preferences from logged interactions. Although some methods operate without unbiased supervision, many studies have shown that even a small amount of unbiased interaction data can substantially improve preference estimation~\cite{AutoDebias, liCausalRecommendationMachine2025}, motivating debiasing frameworks based on meta-learning~\cite{AutoDebias}, knowledge distillation~\cite{liuGeneralKnowledgeDistillation2020, InterD}, and machine unlearning~\cite{liCausalRecommendationMachine2025}. Despite their strong offline performance, existing unbiased recommendation methods still face important challenges in industrial deployment, as they mainly optimize unbiased evaluation metrics while overlooking recommendation quality in factual biased serving environments, where excessive debiasing may harm online engagement and platform revenue. InterD~\cite{InterD} is the first framework to explicitly optimize performance in both biased and unbiased environments through dual-teacher distillation. However, it relies on AutoDebias-based meta-learning~\cite{AutoDebias} and imputation distillation over the full user--item Cartesian product, which incurs expensive bi-level optimization, substantial computational overhead, and limited compatibility with large-scale industrial streaming recommender systems.

To address these challenges, we propose ConAlign (Conditional Alignment Framework), a practical debiasing framework designed for industrial recommender systems. ConAlign employs a unified dual-tower architecture consisting of a biased tower trained on large-scale biased interaction data and an unbiased tower trained on limited unbiased data. The predictions generated by the unbiased tower are used as the final recommendation outputs. Instead of performing unconditional global correction, ConAlign introduces a conditional cross-tower alignment mechanism that selectively transfers knowledge from the biased tower to the unbiased tower, enabling effective debiasing while preserving recommendation accuracy in factual biased environments. The proposed framework supports lightweight online streaming training without requiring expensive Cartesian-product modeling or multi-stage distillation pipelines~\cite{AutoDebias, InterD}. To the best of our knowledge, ConAlign is the first streaming debiasing recommendation framework successfully deployed in a large-scale industrial recommendation system that utilizes a small fraction of unbiased random traffic for debiasing. Extensive offline experiments and real-world online A/B testing on Kuaishou demonstrate that ConAlign consistently improves both unbiased recommendation quality and factual online performance while enhancing long-term user engagement and interest diversity.

In summary, this paper makes the following contributions:

\begin{itemize}[leftmargin=*]

    \item We identify the dual-environment optimization challenge in industrial recommender systems, where recommendation models must simultaneously capture users' intrinsic preferences under unbiased distributions while maintaining strong performance in factual recommendation environments.

    \item We propose ConAlign, a conditional cross-tower distribution alignment framework that selectively transfers knowledge between biased and unbiased representations, effectively improving unbiased recommendation performance while preserving factual accuracy in biased environments.

    \item We successfully deploy the ConAlign framework in a real-world industrial recommendation system, where it supports efficient online streaming training on continuously arriving data.
    
    \item We validate ConAlign through extensive offline experiments and large-scale online A/B testing on Kuaishou's production platform, demonstrating consistent improvements in recommendation accuracy, long-term retention, and interest diversity.

\end{itemize}

\section{Related Work}

\subsection{Debiasing in recommender systems}

Accurately measuring user interests remains a critical and long-standing challenge in recommender systems. Practical recommender systems are typically trained on observational user feedback, which is inevitably affected by various biases, including selection bias~\cite{marlinCollaborativeFilteringMissing2012}, exposure bias~\cite{liangModelingUserExposure2016}, popularity bias~\cite{krishnanAdversarialApproachImprove2018}, and position bias~\cite{collinsStudyPositionBias2018,chenBiasDebiasRecommender2023}. To mitigate the detrimental effects of these biases, numerous debiasing methods have been proposed. Early studies addressed missing feedback through imputation~\cite{steckEvaluationRecommendationsRatingprediction2013,saitoUnbiasedRecommenderLearning2020}. Counterfactual learning methods, including IPS~\cite{IPS}, DR~\cite{wangDoublyRobustJoint2019}, and MRDR~\cite{guoEnhancedDoublyRobust2021}, correct observational bias via propensity estimation and counterfactual estimation~\cite{liRelaxingAccurateImputation2024,liTDRCLTargetedDoubly2023}. More recent approaches further explore invariant learning~\cite{wangInvariantPreferenceLearning2022,KD-Debias}, adversarial learning~\cite{zhangGeneralDebiasingGraphbased2024a}, and other learning paradigms to improve recommendation debiasing.

Nevertheless, relying solely on biased observational data is insufficient to fully uncover users' diverse interests and effectively break the filter bubble~\cite{rowlandFilterBubbleWhat2011}. A substantial body of literature has demonstrated that incorporating a small amount of unbiased data can significantly enhance debiasing performance~\cite{AutoDebias,liCausalRecommendationMachine2025}. For instance, Chen et al.~\cite{AutoDebias} proposed a unified debiasing framework leveraging unbiased data via meta-learning. Similarly, KDCRec~\cite{liuGeneralKnowledgeDistillation2020} and InterD~\cite{InterD} train unbiased teacher models on unbiased data and perform knowledge distillation to debias the student model. Wang et al.~\cite{wangCombatingSelectionBiases2021} employed bi-level optimization to learn propensity scores, and Liu et al.~\cite{liuRatingDistributionCalibration2022} introduced a self-supervised learning paradigm to calibrate rating distributions. Other parallel lines of work attempt to mitigate biases through causal intervention ~\cite{zhangDebiasingRecommendationLearning2023a} or counterfactual inference ~\cite{weiModelAgnosticCounterfactualReasoning2021}. For example, iDCF~\cite{zhangDebiasingRecommendationLearning2023a} applies proximal causal inference to infer unmeasured confounders and identify counterfactual feedback.

Despite these advances, most existing methods primarily optimize recommendation performance in unbiased environments. Such optimization may sacrifice recommendation quality in real-world production environments~\cite{InterD}. InterD was the pioneer in simultaneously considering model performance in both biased and unbiased environments. Nevertheless, it suffers from severe limitations such as low training efficiency and poor operability, rendering its deployment in real-world industrial systems suboptimal. In contrast, our method is designed for practical large-scale recommender systems, effectively leveraging limited unbiased data while preserving recommendation performance in biased online environments.

\subsection{Knowledge Distillation}

Knowledge distillation~\cite{hintonDistillingKnowledgeNeural2015} has been widely adopted in deep learning research in recent years. It first trains a teacher network and then extracts the knowledge encoded in the teacher model as soft labels to guide the training of the final student model. Benefiting from this paradigm, the student model can achieve competitive performance with a simpler architecture and lower inference latency. Knowledge distillation has also been extensively applied in recommender systems, including unbiased recommendation tasks~\cite{liuGeneralKnowledgeDistillation2020, InterD, KD-Debias}. Among previous debiasing studies based on knowledge distillation, KDCRec~\cite{liuGeneralKnowledgeDistillation2020} proposes four distillation paradigms that leverage a small amount of unbiased data to correct the bias of the student model. However, since the unbiased model is directly used as the teacher model, the limited amount of unbiased data may constrain both the teacher model quality and the final estimation accuracy of the student model. InterD~\cite{InterD} further transfers knowledge from both biased and unbiased teacher models, enabling the student model to achieve competitive performance in both environments. KD-Debias~\cite{KD-Debias} demonstrates that knowledge distillation can also achieve strong unbiased recommendation performance without relying on unbiased data. Nevertheless, both InterD~\cite{InterD} and KD-Debias~\cite{KD-Debias} only perform knowledge distillation at the final prediction-score level. Inspired by the success of knowledge distillation, our method adopts a more flexible cross-level distillation strategy, enabling knowledge transfer beyond the final prediction layer.

\section{METHODOLOGY}

\begin{figure*}[h]
  \centering
  \includegraphics[width=0.75\linewidth]{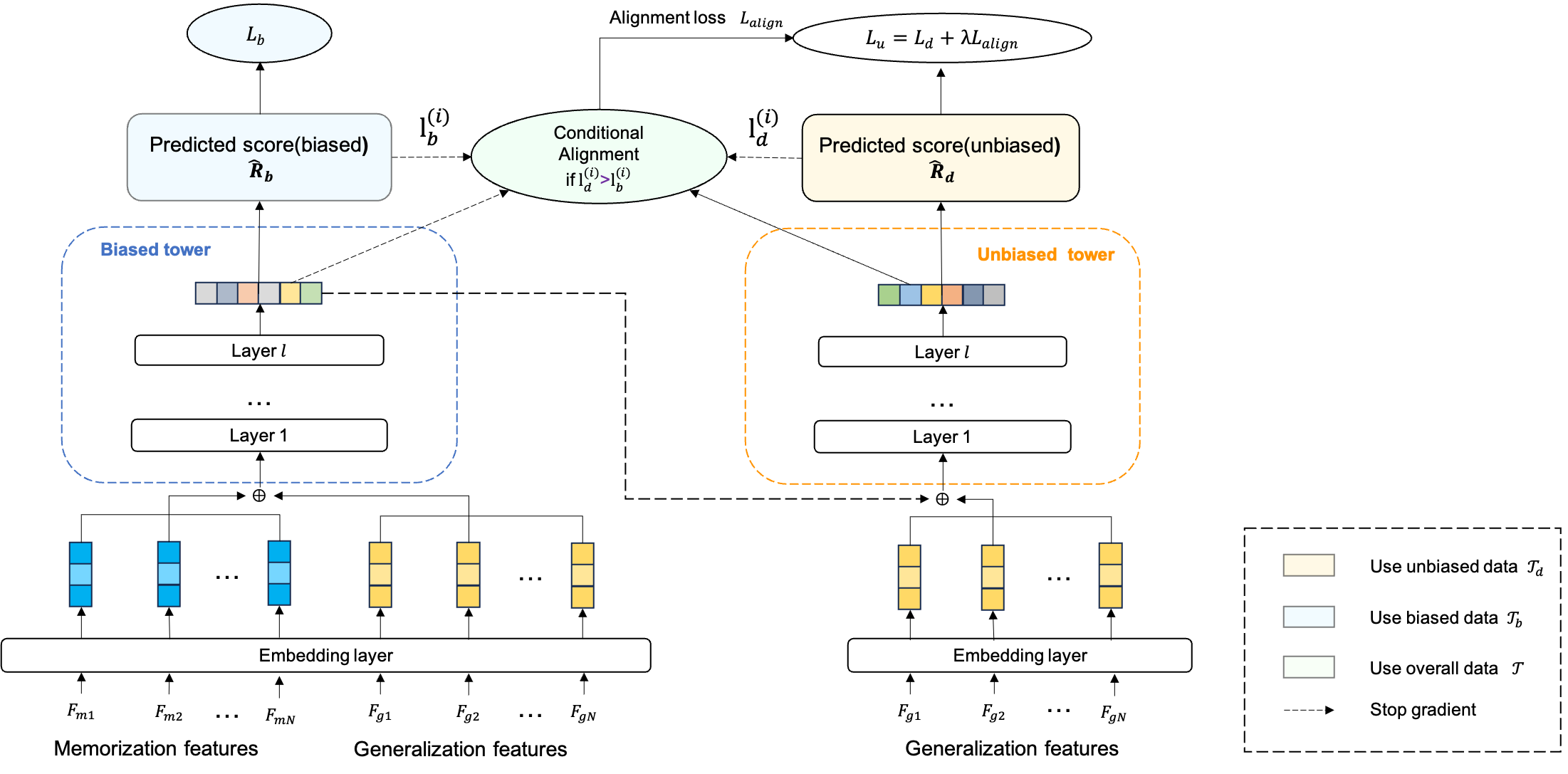}
  \caption{Overview of the ConAlign framework. ConAlign consists of a biased tower trained on biased interaction data and an unbiased tower trained on unbiased data. A conditional alignment loss is introduced to jointly preserve recommendation performance in both biased and unbiased environments.}
  \label{fig:model}
\end{figure*}

\subsection{Problem Formulation}
An ideal recommender system aims to estimate users' intrinsic preferences over the entire user--item space. Let $\mathcal{U}$ and $\mathcal{I}$ respectively denote the user set and item set, and define the complete interaction space as $\mathcal{D}=\mathcal{U}\times\mathcal{I}$. For each interaction pair $(u,i)\in\mathcal{D}$, let $R(u,i)$ denote the latent user preference, which may correspond to various feedback signals such as click-through rate (CTR), favorite rate, watch time, or explicit ratings. The ultimate recommendation objective is therefore to estimate the unbiased preference distribution $P(R|\mathcal{D})$.

However, real-world recommender systems are primarily trained on logged observational feedback rather than uniformly sampled interactions. Let $O(u,i)\in\{0,1\}$ denote the exposure variable, where $O(u,i)=1$ indicates that item $i$ is exposed to user $u$. Since user feedback can only be collected on exposed items, the observed interaction data are actually sampled from $P(\mathcal{D}_b)=P(\mathcal{D}|O=1)$ instead of the ideal distribution $P(\mathcal{D})$. Consequently, the empirical interaction space $\mathcal{D}_b\subseteq\mathcal{D}$ is inherently biased.

Moreover, the exposure mechanism is affected by multiple confounding factors, including historical recommendation policies, item popularity, position bias, and user activity bias. Let $e$ denote the aggregated environmental effect induced by these biases. In practice, recommendation models are optimized on the biased feedback distribution $P(R_e|\mathcal{D}_b)$ rather than the target unbiased distribution $P(R|\mathcal{D})$.

In practical recommendation scenarios, modeling both biased and unbiased preference distributions is equally important. Estimating $P(R|\mathcal{D})$ helps capture users' intrinsic interests and improves unbiased recommendation quality, while modeling $P(R_e|\mathcal{D}_b)$ remains critical for maintaining recommendation accuracy in factual recommendation environments. More importantly, jointly considering these two distributions can alleviate the filter bubble phenomenon and improve the long-term diversity and sustainability of recommendation ecosystems.

In this work, we leverage a large-scale biased observational dataset sampled from $D_b$ together with a limited amount of unbiased interaction data sampled from $D$ to learn a recommendation model that achieves strong performance under unbiased environments while maintaining competitive performance in biased factual environments.

\subsection{Model Architecture}

Figure \ref{fig:model} illustrates the overall architecture of the Conditional Alignment Framework (ConAlign), which is built upon a unified dual-tower~\cite{yiSamplingbiascorrectedNeuralModeling2019} design. The framework consists of a biased tower trained on biased interaction data and an unbiased tower trained on unbiased interaction data, where distribution alignment between the two towers is achieved through an alignment loss. During inference, the prediction score generated by the unbiased tower is used as the final recommendation output.

\subsubsection{Cross-tower Architecture}

As illustrated in Figure \ref{fig:model}, ConAlign encompasses both a biased model and an unbiased model architecture. The input of the biased tower contains the complete feature information of users, items, and their interactions. These features consist of memorization features and generalization features, formulated as:
\begin{equation}
    \mathbf{x}^{(b)} =[E(F_m) \oplus E(F_g)]
\end{equation}
where $\oplus$ denotes the concatenation operation. $F_m$ represents the memorization features, which typically capture the historical combinational information of users and items. $F_g$ represents the generalization features, encompassing inherent attributes of users and items, such as profiles, categories, and contextual information.

The biased tower is trained exclusively on the biased interaction dataset $\mathcal{T}_b=\{(u,i,r_b)\}$, where $(u,i)\in\mathcal{D}_b$ denotes an observed user--item interaction pair sampled from the biased interaction space, and $r_b$ represents the corresponding biased feedback label, such as click, watch, or rating signals collected under the factual recommendation environment. The biased input $\mathbf{x}^{(b)}$ is fed into the biased tower to generate the predicted score $\hat{R}_b=f_b(\mathbf{x}^{(b)})$. The biased tower is optimized by minimizing the following empirical loss:
\begin{equation}
    \mathcal{L}_b
    =
    \frac{1}{|\mathcal{T}_b|}
    \sum_{(u,i,r_b)\in\mathcal{T}_b}
    \ell(\hat{R}_b(u,i), r_b)
\end{equation}

Unlike conventional independent two-tower architectures, our framework adopts a cross-tower design that transfers latent representations from the biased tower to the unbiased tower. This mechanism enables the unbiased tower to leverage semantic information learned from large-scale biased interactions, thereby improving representation quality and stabilizing unbiased learning. Specifically, we extract the last hidden representation of the biased tower as the latent bias representation (LBR), and concatenate it with the invariant generalization features to construct the input of the unbiased tower:
\begin{equation}
    \mathbf{x}^{(d)} = \mathbf{h}_b^{(L-1)} \oplus E(F_g)
\end{equation}
where $\oplus$ denotes feature concatenation, $L$ is the number of layers in the biased tower, and $\mathbf{h}_b^{(L-1)}$ denotes the last hidden representation of the biased tower, since the final layer is used to generate the predicted recommendation score.

The unbiased tower is trained using unbiased interaction data $\mathcal{T}_d=\{(u,i,R_d)\}$, where $R_d$ denotes the unbiased feedback label associated with user--item pair $(u,i)$. The unbiased prediction is formulated as
\begin{equation}
    \hat{R}_d = f_d(\mathbf{x}^{(d)}),
\end{equation}
and the unbiased optimization objective is defined as
\begin{equation}
    \mathcal{L}_d
    =
    \frac{1}{|\mathcal{T}_d|}
    \sum_{(u,i,R_d)\in\mathcal{T}_d}
    \ell(\hat{R}_d(u,i),r_d),
\end{equation}
where $\ell(\cdot)$ denotes the base loss function, such as BCE loss. Crucially, during unbiased training, gradients are prevented from propagating back to the biased tower, ensuring that unbiased optimization does not interfere with the modeling capability of the biased tower.

\subsubsection{Conditional Alignment Loss}

While the unbiased tower is explicitly guided by unbiased data, optimizing it in pure isolation often leads to degradation in factual accuracy due to environmental discrepancy. To successfully balance unbiased generalizability and factual accuracy, we introduce a cross-layer Conditional Alignment Loss to regularize the latent space of the unbiased tower:
\begin{equation}
    \mathcal{L}_{align} = \mathbb{I}_{cond} \cdot \frac{1}{N} \sum \left| \mathbf{h}_d^{(L-1)} - \text{sg}(\mathbf{h}_b^{(L-1)}) \right|
\end{equation}
where $\mathbf{h}_b^{(L-1)}$ and $\mathbf{h}_d^{(L-1)}$ denote the penultimate representations of the biased and unbiased modules, i.e., before the final prediction layer. We perform alignment at the latent representation level rather than the final prediction-score level, enabling the unbiased tower to distill richer collaborative capacity from the biased tower without just memorizing final biased probabilities. The effectiveness of this design will be further validated in Section~\ref{tab:Ablation Study}.The operation $\text{sg}(\cdot)$ stands for \textit{stop-gradient}, ensuring that the gradients from the alignment loss do not flow back to update the biased tower.

To avoid negative transfer and prevent the unbiased tower from being excessively regularized toward factual biases, we do not impose the alignment objective uniformly across all training instances. Instead, we introduce a conditional alignment mechanism, where knowledge transfer is activated only when the biased tower provides more reliable supervision than the unbiased tower. Concretely, for the $i$-th training instance in a mixed mini-batch, we define a gating indicator as
\begin{equation}
\mathbb{I}_{\mathrm{cond}}^{(i)}=
\begin{cases}
1, & \text{if } \ell_{\mathrm{b}}^{(i)} < \ell_{\mathrm{d}}^{(i)},\\
0, & \text{otherwise},
\end{cases}
\end{equation}
where $\ell_{\mathrm{b}}^{(i)}$ and $\ell_{\mathrm{d}}^{(i)}$ denote the prediction losses of the biased tower and the unbiased tower, respectively, on the $i$-th instance. The mixed mini-batch consists of both biased and unbiased interactions, and the alignment term is activated for any interaction on which the biased tower incurs a lower prediction loss than the unbiased tower. In this way, the unbiased tower can benefit from factual signals where necessary, while retaining sufficient flexibility to capture unbiased preference patterns.

Meanwhile, the prediction objectives of the two towers are still optimized on their respective supervision sources. In particular, $\mathcal{L}_{\mathrm{b}}$ and $\mathcal{L}_{\mathrm{d}}$ are computed on their corresponding data subsets, while the alignment term is selectively accumulated over the instances satisfying $\mathbb{I}_{\mathrm{cond}}^{(i)}=1$.

By incorporating this conditional constraint, the total objective function $\mathcal{L}_{u}$ for the unbiased tower is formulated as:
\begin{equation}
    \mathcal{L}_{u} = \mathcal{L}_d + \lambda \mathcal{L}_{align}
\end{equation}
where $\lambda$ is a hyperparameter balancing unbiased supervision and factual alignment. 

During inference, ConAlign directly deploys the prediction score $\hat{R}_d$ as the final recommendation output.

\subsection{Online Implementation Details}

\subsubsection{Unbiased Data Collection}

Standard exposure logs are entangled with the system's own recommendation policy, making it impossible to recover true user preferences from biased data alone. A principled remedy is to collect \emph{missing-at-random} (MAR) data via random intervention~\cite{gaoKuaiRandUnbiasedSequential2022}.

Concretely, each time the recommendation system generates a ranked list, we trigger a random intervention with a fixed probability. When triggered, one video is uniformly sampled from the full candidate pool and inserted at a uniformly random position in the returned feed, bypassing all ranking and personalization stages. Because both item selection and insertion position are uniform and independent of the system policy, the resulting interaction signals satisfy the missing-at-random (MAR) condition. Interactions on these randomly inserted items are tagged and collected into the unbiased stream $\mathcal{D}_d$, kept strictly separate from the main biased log $\mathcal{D}_b$. The intervention probability is deliberately kept small, so that randomly inserted items constitute only a negligible fraction of 
the overall feed, ensuring minimal impact on short-term user engagement.

Unlike KuaiRand~\cite{gaoKuaiRandUnbiasedSequential2022}, which conducted random interventions over a fixed two-week window to release a static research dataset, our unbiased traffic stream is a \emph{permanent, continuously running} component of the production system. This allows the unbiased tower to be trained in an online streaming fashion, adapting to evolving user interests and item distributions in real time, which is essential for deployment in a dynamic industrial environment.

\subsubsection{Streaming Training}

ConAlign is trained in an online streaming fashion on two concurrent data streams. Due to the large volume disparity between the two streams, each training batch is constructed by sampling a fixed number of examples from each stream independently. The biased samples update the biased tower exclusively, while the unbiased samples drive the unbiased tower. The adaptive alignment condition $\mathbb{I}_{cond}$ is evaluated within each batch before the backward pass, and gradients from the alignment loss are applied only to the unbiased tower via the stop-gradient operation on $\mathbf{h}_b^{(L-1)}$.

\subsubsection{Online Serving}

At inference time, only the prediction score generated by the unbiased tower, $\hat{R}_d$, is used as the final output for ranking. This is made feasible by the alignment loss: by regularizing the unbiased tower toward the biased tower when the former underperforms, ConAlign ensures the unbiased tower maintains competitive accuracy in the biased online environment while still capturing users' true preferences. Since the unbiased tower is a lightweight network built on top of the biased tower's representation, it introduces only negligible additional latency to the standard recommendation pipeline.

\section{Experiments}

In this section, we conduct extensive experiments to evaluate the effectiveness of ConAlign and answer the following research questions:

\begin{itemize}
    \item \textbf{RQ1}: Compared with existing debiasing methods, can ConAlign achieve better accuracy in unbiased environments while preserving performance in biased factual environments?
    
    \item \textbf{RQ2}: What are the contributions of different components in ConAlign?
    
    \item \textbf{RQ3}: How does ConAlign perform in real-world industrial deployment scenarios?
\end{itemize}

\subsection{Offline Experimental Settings}

\begin{table*}[htbp]
\centering
\caption{Performance comparison on three datasets under biased and unbiased test environments. The best results are highlighted in \textbf{bold}. The last column reports the training latency on Yahoo! R3 (in seconds).}
\label{tab:main_results}

\resizebox{\textwidth}{!}{
\renewcommand{\arraystretch}{1.2}
\begin{tabular}{l c c c c | c c c c | c c c c | c}
\hline \hline

\multicolumn{1}{c}{Dataset} 
& \multicolumn{4}{c|}{Coat} 
& \multicolumn{4}{c|}{Yahoo! R3} 
& \multicolumn{4}{c|}{KuaiRand-Pure}
& \multirow{2}{*}{\begin{tabular}[c]{@{}c@{}}Yahoo! R3\\ Time (s)\end{tabular}} \\
\cline{1-13}

\multirow{2}{*}{Method} 
& \multicolumn{2}{c}{Biased} 
& \multicolumn{2}{c|}{Unbiased} 
& \multicolumn{2}{c}{Biased} 
& \multicolumn{2}{c|}{Unbiased} 
& \multicolumn{2}{c}{Biased} 
& \multicolumn{2}{c|}{Unbiased} 
&  \\
\cline{2-13}

& UAUC & NDCG 
& UAUC & NDCG 
& UAUC & NDCG 
& UAUC & NDCG 
& UAUC & NDCG 
& UAUC & NDCG 
&  \\
\hline

MF  
& 0.6754 & 0.7611 & 0.6653 & 0.6844 
& 0.6569 & 0.7910 & 0.6456 & 0.7455 
& 0.6109 & 0.8019 & 0.5622 & 0.5001 
& 52.6 \\

MF-IPS   
& 0.6752 & 0.7609 & 0.6744 & 0.6924 
& 0.6498 & 0.7882 & 0.6444 & 0.7433 
& 0.5975 & 0.7961 & 0.5418 & 0.4791 
& 54.2 \\

MF-DR    
& 0.6728 & 0.7520 & 0.6836 & 0.6970
& 0.6494 & 0.7878 & 0.6454 & 0.7443 
& 0.5627 & 0.7808 & 0.4664 & 0.4153  
& 88.1 \\

CausE 
& 0.6744 & 0.7603 & 0.6687 & 0.6863 
& 0.6511 & 0.7902 & 0.6537 & 0.7505
& 0.5672 & 0.7813 & 0.5422 & 0.4913 
& 126.4 \\

KD-label 
& 0.6903 & 0.7692 & 0.6683 & 0.6860 
& 0.6428 & 0.7843 & 0.6608 & 0.7577 
& 0.6049 & 0.7989 & 0.6010 & 0.5392 
& 247.7 \\

AutoDebias 
& 0.6658 & 0.7596 & 0.6637 & 0.6744 
& 0.6343 & 0.7804 & 0.6679 & 0.7654 
& 0.6012 & 0.7967 & 0.5958 & 0.5376 
& 4560.7 \\

InterD 
& 0.6733 & 0.7595 & 0.6792 & 0.6939 
& \textbf{0.6622} & \textbf{0.7938} & 0.6787 & 0.7713 
& \textbf{0.6190} & \textbf{0.8057} & 0.6251 & 0.5544  
& 10592.1 \\

KD-debias 
& 0.6098 & 0.7197 & 0.5961 & 0.6369 
& 0.6472 & 0.7870 & 0.6523 & 0.7517 
& 0.6083 & 0.7997 & 0.5675 & 0.5133 
& 466.9 \\
\hline

\textbf{Ours Model} 
& \textbf{0.6954} & \textbf{0.7746} 
& \textbf{0.6856} & \textbf{0.7066} 
& 0.6535 & 0.7898 
& \textbf{0.6841} & \textbf{0.7761} 
& 0.6145 & 0.8039 
& \textbf{0.6372} & \textbf{0.5656} 
& 122.7 \\

\hline \hline
\end{tabular}
}
\end{table*}

\subsubsection{Datasets}

To comprehensively evaluate the effectiveness of ConAlign, we conduct experiments on three widely used recommendation datasets containing unbiased interaction data.

\paragraph{Coat}
The Coat dataset simulates users purchasing coats in an online shopping platform. It contains 6,960 biased ratings from 290 users on 300 items, where each user selectively rated 24 items according to personal preferences. In addition, the dataset~\cite{IPS} provides 4,640 unbiased ratings, where each user was required to rate 16 randomly selected items.

\paragraph{Yahoo! R3}
Yahoo! R3 is a music recommendation dataset~\cite{marlinCollaborativePredictionRanking2009} containing user ratings on songs. The biased subset contains 311,704 ratings from 15,400 users on self-selected songs, while the unbiased subset contains ratings from 5,400 users on 10 randomly selected songs.

\paragraph{KuaiRand-Pure}
KuaiRand is an unbiased sequential recommendation dataset~\cite{gaoKuaiRandUnbiasedSequential2022} collected from the real-world recommendation logs of the Kuaishou short-video platform. It is the first recommendation dataset that introduces millions of randomly exposed intervention items into a standard recommendation feed. KuaiRand-Pure contains both biased and unbiased interaction data between 1,000 users and 7,388 videos.

Following the experimental settings of ~\cite{InterD}, for the Coat and Yahoo! R3 datasets, ratings larger than 3 (on a 1--5 scale) are treated as positive samples, while the remaining ratings are regarded as negative samples. We randomly sample 70\% of the biased data and 50\% of the unbiased data as the training set. In addition, 10\% of the biased and unbiased data are used as the biased validation set and unbiased validation set, respectively. The remaining 20\% biased data and 40\% unbiased data are used for testing.

For KuaiRand-Pure, we use interaction logs from April 22 to May 8, covering 17 consecutive days, and model the binary \texttt{is\_click} field. In the two-column UI, it indicates a click; in the single-column UI, it means valid\_play: which equals 1 when: play\_time\_ms >= duration\_ms if duration\_ms <= 7,000 ms, or play\_time\_ms > 7,000 ms if duration\_ms > 7,000 ms. The biased and unbiased data from the first two weeks are used for training. The interactions on Day 15 are used for validation, while the last two days are used for testing.

\subsubsection{Baselines}

We compare ConAlign with several representative debiasing methods in recommender systems.

\begin{itemize}
    \item \textbf{MF}~\cite{korenMatrixFactorizationTechniques2009}: Standard matrix factorization model.
    
    \item \textbf{MF-IPS}~\cite{IPS}: A causal debiasing method that reweights the loss function using inverse propensity scores. In our experiments, the propensity scores are estimated using Bayesian estimation.
    
    \item \textbf{MF-DR}~\cite{wangDoublyRobustJoint2019}: A doubly robust estimator that combines inverse propensity weighting and imputation to improve both accuracy and robustness.
    
    \item \textbf{CauseE}~\cite{liuGeneralKnowledgeDistillation2020}: A debiasing framework that guides unbiased representation learning through embedding-level contrastive learning.
    
    \item \textbf{KD-label}~\cite{liuGeneralKnowledgeDistillation2020}: A knowledge distillation method that uses a model trained on unbiased data as the teacher model. The teacher predictions are used as soft labels to train the student model. We adopt the best label-based distillation strategy reported in the original paper.
    
    \item \textbf{AutoDebias}~\cite{AutoDebias}: A unified meta-learning-based debiasing framework.
    
    \item \textbf{InterD}~\cite{InterD}: The current state-of-the-art method that simultaneously considers performance in both biased and unbiased environments. It employs MF and AutoDebias as biased and unbiased teacher models, respectively, and trains the final student model via knowledge distillation.
    
    \item \textbf{KD-Debias}~\cite{KD-Debias}: A state-of-the-art debiasing method using only biased data. It leverages invariant learning and distance-aware knowledge distillation for unbiased model training.
\end{itemize}

\subsubsection{Evaluation Metrics and Implementation Details}

We adopt two widely used recommendation metrics, UAUC and NDCG@5, to evaluate model performance. For public benchmark experiments, Matrix Factorization (MF)~\cite{korenMatrixFactorizationTechniques2009} is selected as the common backbone for all methods, following prior debiasing studies and ensuring fair comparison with baselines such as AutoDebias and InterD. This benchmark setting does not restrict the ConAlign framework itself, which only relies on latent representations and prediction scores from the two towers.

For each compared method, we conduct grid search to determine the optimal hyperparameter configuration. The search space for the learning rate and weight decay is \{1, 0.1, 0.01, 0.001, 0.00001, 0.000001\}

\subsection{Offline Overall Performance (RQ1)}

Table~\ref{tab:main_results} reports the experimental results of all methods under both biased and unbiased testing environments on the three datasets. From the results, we have the following observations.

First, we compare representative debiasing methods that utilize unbiased data (i.e., AutoDebias and InterD) with the method that does not leverage unbiased data (i.e., KD-Debias). The results indicate that methods incorporating unbiased data generally achieve better performance under unbiased evaluation settings. This shows that properly exploiting unbiased data can effectively capture users’ latent true preferences and further enhance debiasing performance. 

Second, \textbf{ConAlign consistently outperforms all baselines in unbiased recommendation performance across all datasets}. These results demonstrate that ConAlign can better capture users' true interests under unbiased environments, leading to more effective recommendation performance in practical applications.

Third, under unbiased evaluation settings, most debiasing methods achieve improvements over the basic MF model. However, in biased factual environments, most debiasing methods suffer from noticeable performance degradation. This observation validates our assumption in Section~3.1 regarding the discrepancy between biased and unbiased environments, and further highlights the necessity of simultaneously optimizing performance in both scenarios. In contrast, ConAlign maintains competitive performance in biased environments while achieving strong debiasing capability. Compared with the current state-of-the-art model InterD, \textbf{ConAlign achieves comparable or superior testing performance across different datasets in biased environments}.

Finally, we further evaluate the training efficiency of different methods on the Yahoo! R3 training set under a consistent experimental setting with 102 CPU cores, 495 GiB memory, and 2 GPUs. As shown in Table~\ref{tab:main_results}, most conventional methods, including MF, MF-IPS, MF-DR, CausE, and KD-label, exhibit relatively similar training latency, where the differences mainly arise from variations in optimization steps and training epochs. In contrast, AutoDebias introduces substantially higher computational overhead due to its meta-learning-based bi-level optimization. InterD further increases the training cost by employing MF and AutoDebias as dual teacher models and performing imputation-based distillation over the full user--item interaction space, resulting in substantially higher computational overhead. Despite achieving competitive recommendation performance, ConAlign maintains a lightweight training profile with significantly lower latency than AutoDebias and InterD, demonstrating better scalability and practical suitability for large-scale industrial recommender systems.

\subsection{In-depth Analyses (RQ2)}

\subsubsection{Ablation Study}
\label{tab:Ablation Study}
To investigate the contributions of different components in ConAlign, we conduct ablation studies on the Yahoo! R3 dataset. Specifically, we construct the following model variants:

\begin{itemize}
    \item \textbf{ours-biasTower}: Only the biased tower of ConAlign is retained.
    
    \item \textbf{ours-noAlign}: The alignment loss is removed during the training of the unbiased tower.
    
    \item \textbf{ours-allAlign}: The adaptive alignment condition used to regulate the alignment loss is removed during unbiased tower training.
    \item \textbf{ours-score}: During the training of the unbiased tower in ConAlign, the alignment loss uses final prediction-score alignment instead of latent representation alignment
\end{itemize}

The experimental results are shown in Table~\ref{tab:ablation}. As shown in Table~\ref{tab:ablation}, compared with \textit{ours-biasTower}, ConAlign achieves substantial improvements in the unbiased evaluation environment during unbiased data training, while maintaining comparable performance in the biased evaluation environment. Furthermore, compared with all other ablation variants, ConAlign consistently achieves superior performance across both biased and unbiased evaluation settings, demonstrating the effectiveness of each proposed component under dual-environment optimization.

In particular, the alignment loss effectively regularizes the unbiased learning process through biased latent representations, which not only alleviates performance degradation in the biased environment but also provides richer semantic information for unbiased preference learning, thereby improving unbiased recommendation performance. The adaptive alignment condition enables the model to utilize biased representations in a more selective and effective manner, avoiding excessive regularization from biased signals. Compared with final prediction-score alignment, latent representation alignment preserves substantially richer structural and semantic information about user preferences, allowing the unbiased tower to learn more informative and transferable representations.

\begin{table}[t]
\centering
\caption{Ablation study of ConAlign on Yahoo! R3 under biased and unbiased evaluation environments.}
\label{tab:ablation}
\small
\begin{tabular}{lcccc}
\toprule
& \multicolumn{2}{c}{\textbf{Biased Test Set}}
& \multicolumn{2}{c}{\textbf{Unbiased Test Set}} \\
\cmidrule(lr){2-3} \cmidrule(lr){4-5}
\textbf{Variant}
& \textbf{UAUC}
& \textbf{NDCG@5}
& \textbf{UAUC}
& \textbf{NDCG@5} \\
\midrule
ours-biasTower        & \textbf{0.6569} & \textbf{0.7910} & 0.6456          & 0.7455          \\
ours-noAlign          & 0.6294          & 0.7774          & 0.6834          & 0.7710         \\
ours-allAlign         & 0.6530          & 0.7903          & 0.6722          & 0.7659          \\
ours-score            & 0.6519          & 0.7893          & 0.6822          & 0.7661          \\
ConAlign                  & 0.6535          & 0.7898          & \textbf{0.6841} & \textbf{0.7761} \\
\bottomrule
\end{tabular}
\end{table}

From the results, removing any component consistently degrades the recommendation performance, demonstrating that each module contributes positively to the final model. 

\subsubsection{Hyperparameter Analysis}

\begin{figure}[h]
  \centering
  \includegraphics[width=0.6\linewidth]{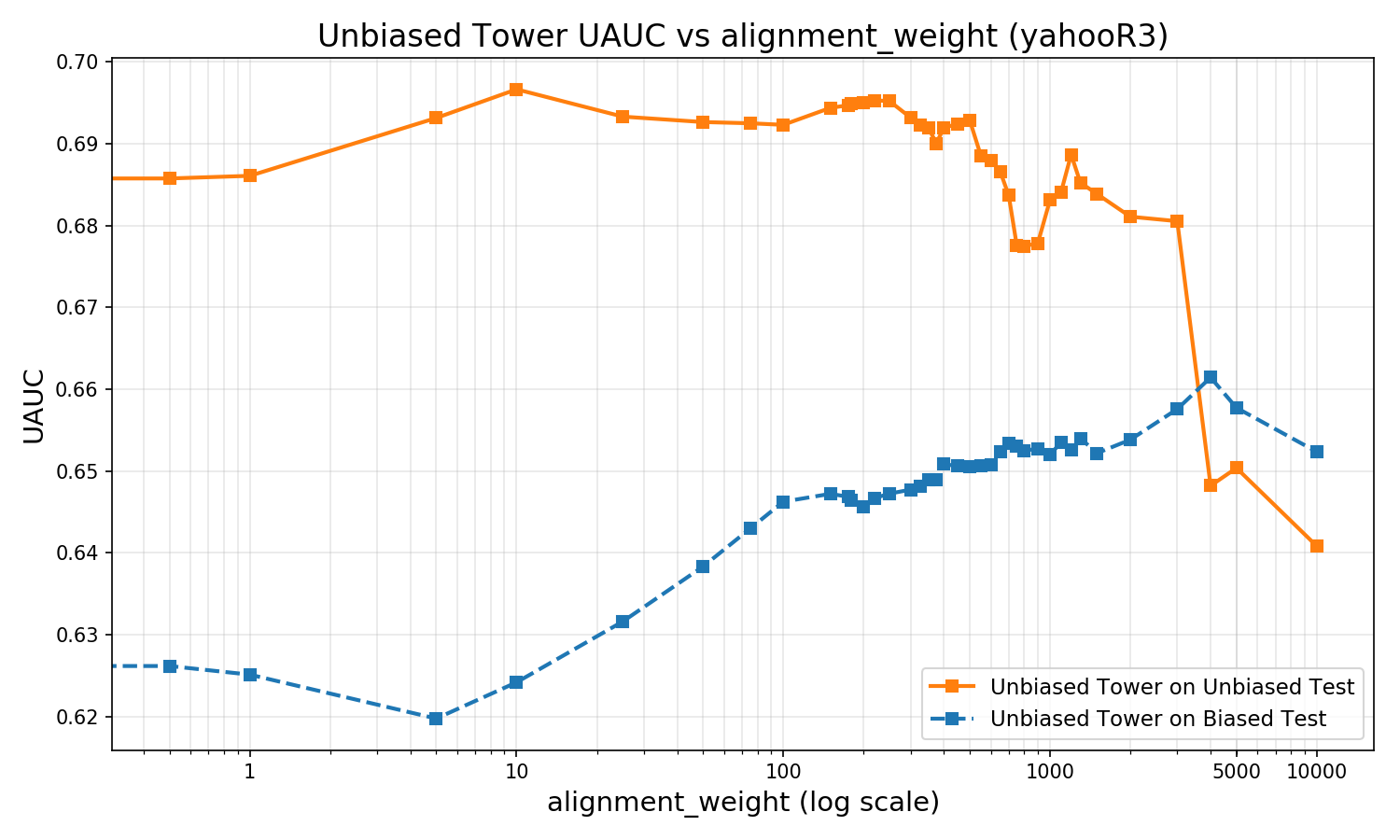}
  \caption{UAUC performance of ConAlign under different alignment weight settings on biased and unbiased test sets.}
  \label{fig:uauc_align}
\end{figure}

\begin{figure}[h]
  \centering
  \includegraphics[width=0.6\linewidth]{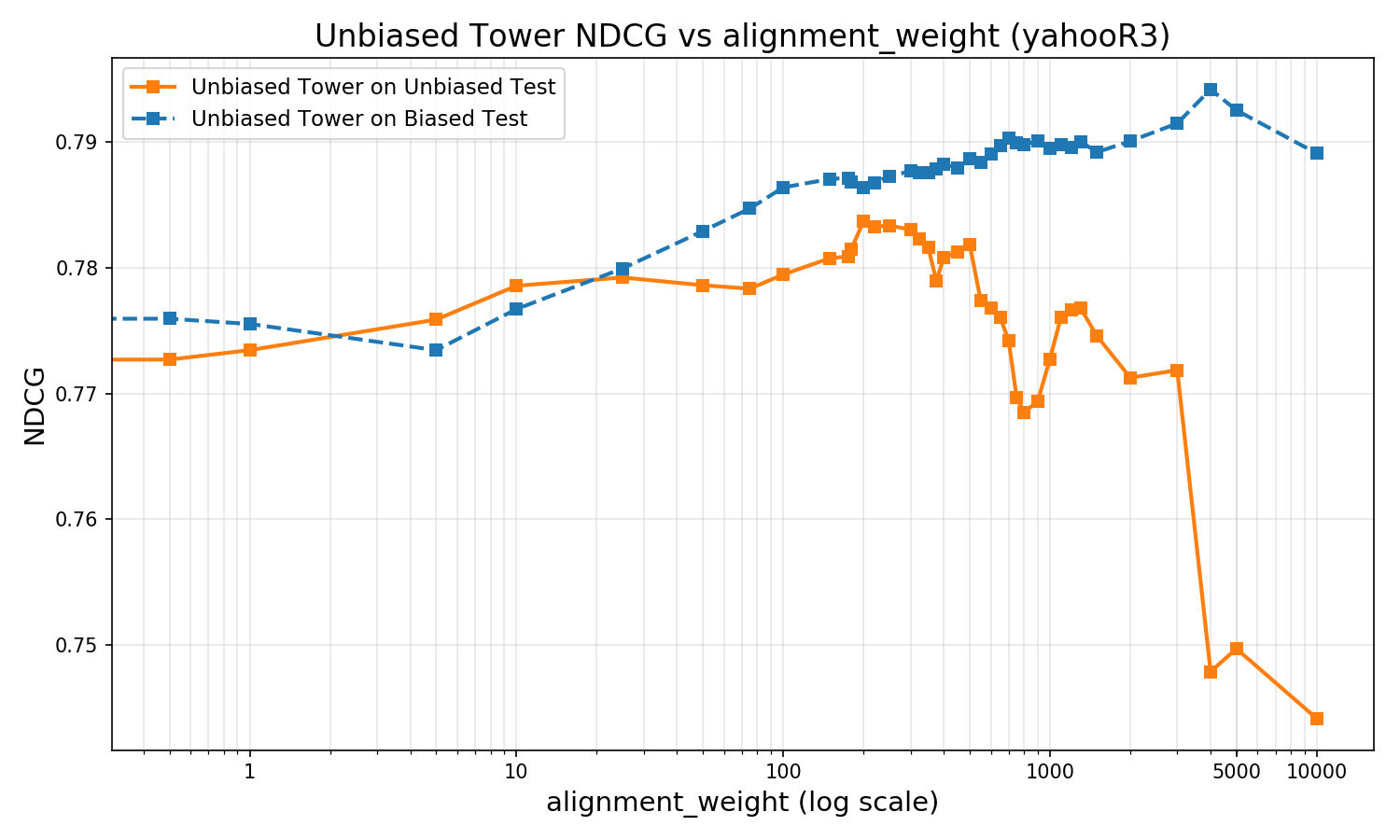}
  \caption{NDCG@5 performance of ConAlign under different alignment weight settings on biased and unbiased test sets.}
  \label{fig:ndcg_align}
\end{figure}

To investigate the impact of the alignment weight on ConAlign, we conduct experiments on the Yahoo! R3 dataset using different parameter settings. The corresponding results are reported in Figure~\ref{fig:uauc_align} and Figure~\ref{fig:ndcg_align}. From the results, we observe that dynamically adjusting the alignment weight effectively controls the trade-off between performance in biased and unbiased environments, demonstrating the flexibility of the proposed framework. Overall, as the alignment weight increases, performance in the biased environment exhibits an upward trend, while performance in the unbiased environment shows a downward trend, which is consistent with the intuition discussed in the Method section. Notably, when the alignment weight becomes excessively large, the performance in the unbiased test environment drops sharply, further highlighting the necessity of preventing the debiased tower from being over-constrained by the biased tower. Nevertheless, compared with the setting where the alignment weight is zero, an appropriate alignment weight can still improve performance in both biased and unbiased scenarios. This observation further corroborates the findings in the ablation study.

\subsection{Live A/B Experiment (RQ3)} 
\label{sec:Live A/B Experiment(RQ3)}

\begin{figure}[t]
\centering
\includegraphics[width=0.7\linewidth]{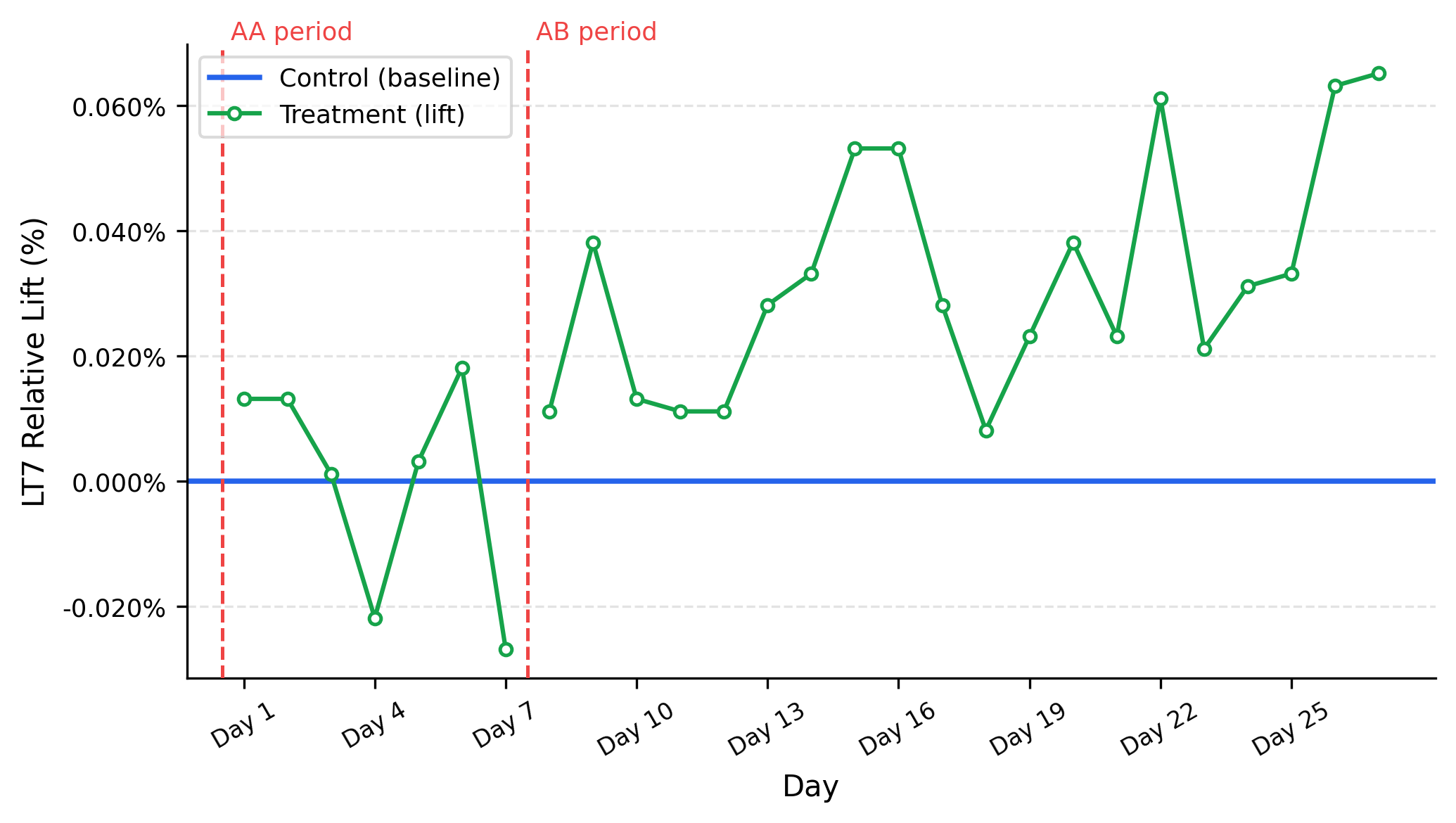}
 \caption{Day-by-day LT7 relative lift over the AA and AB periods. The AA period fluctuates around zero, confirming group balance, while the AB period shows a consistent positive trend.}
\label{fig:ab_results}
\end{figure}

We deployed ConAlign on Kuaishou's production short-video recommendation system, where the two towers are instantiated with the production deep neural ranking backbone rather than the MF backbone used in public benchmark experiments. Before online deployment, ConAlign passed Kuaishou’s standard large-scale offline evaluation on production logs. We do not report the internal offline metrics due to business and data sensitivity. Instead, we report the controlled online A/B results from a 10\% user traffic split below. We first ran a 7-day AA period to confirm group comparability, followed by a long-term AB observation period. Figure~\ref{fig:ab_results} shows the day-by-day LT7 (core retention metric) trend over the AA and AB periods.

Apart from retention, we also track two complementary diversity metrics throughout the experiment: \textbf{Valid Interest Number (VIN)}, the number of interest categories in which a user has accumulated sufficient positive feedback over 7 days; and \textbf{Category Concentration (CC)}, a session-level score computed over a sliding window, where higher values indicate narrower exposure. Table~\ref{tab:online} summarizes the key results with confidence interval (CI).

\begin{table}[h]
\centering
\caption{Online A/B test results of ConAlign on Kuaishou's production platform. All gains are relative improvements. \textbf{Bold} entries indicate statistically significant changes.}
\label{tab:online}
\begin{tabular}{llrr}
\toprule
\textbf{Category} & \textbf{Metric} & \textbf{Gain} & \textbf{95\% CI} \\
\midrule
\multirow{2}{*}{Retention} 
  & \textbf{DAU}             & +0.069\% & [+0.01\%, +0.13\%] \\
  & \textbf{LT7}             & +0.029\% & [+0.005\%, +0.055\%] \\
\midrule
\multirow{2}{*}{Diversity}  
  & \textbf{VIN} & +0.097\% & [+0.01\%, +0.18\%] \\
  & \textbf{CC} & -0.083\% & [-0.15\%, -0.02\%] \\
\midrule
\multirow{2}{*}{Engagement} 
  & \textbf{Completion Plays}  & +0.277\% & [+0.11\%, +0.45\%] \\
  & \textbf{Effective Plays}   & +0.184\% & [+0.03\%, +0.33\%] \\
\midrule
\multirow{2}{*}{Constraint} 
  & Watch Duration    & +0.075\% & [-0.06\%, +0.20\%] \\
  & Rank CPU Cost     & -0.169\% & [-1.02\%, +0.68\%] \\
\bottomrule
\end{tabular}
\end{table}

ConAlign achieves significant gains in both retention and diversity. The improvement in DAU and LT7 is consistent with our motivating analysis: by recovering latent user interests suppressed by biased training, the system reduces filter bubble effects and improves long-term user engagement. Both diversity metrics move in the favorable direction, corroborating that ConAlign broadens content exposure. Short-term engagement metrics: completion plays, effective plays, and watch duration remain positive or statistically neutral, confirming that the diversity improvement does not come at the cost of immediate user experience.

It is worth noting that retention metrics such as DAU and LT7 
are notoriously difficult to improve at scale in mature 
industrial recommender systems, where most algorithmic 
optimizations yield negligible or even negative effects on 
long-term engagement. The observed gains of +0.069\% in DAU 
and +0.029\% in LT7 are therefore practically significant.

\section{Conclusion}
\label{sec:Conclusion}

In this paper, we investigate how to better explore and characterize users' intrinsic preferences in unbiased environments while minimizing performance degradation in factual biased recommendation scenarios. To this end, we propose a cross-tower framework, where the biased and unbiased towers are trained using biased and unbiased interaction data, respectively. Specifically, we introduce an alignment loss that leverages biased samples to guide the learning of unbiased representations, thereby improving the recommendation accuracy in the unbiased space while preventing significant performance loss in the biased space. We conduct extensive offline and online experiments to evaluate the proposed method. Empirical results demonstrate that ConAlign achieves strong performance in both biased and unbiased evaluation environments. Furthermore, in the real-world recommendation scenario of Kuaishou, we design practical engineering strategies and successfully deploy the proposed framework in the production system, leading to consistent improvements across multiple online metrics.

\clearpage

\bibliographystyle{ACM-Reference-Format}
\balance
\bibliography{references1}

@misc{KD-Debias,
  title = {Invariant Debiasing Learning for Recommendation via Biased Imputation},
  author = {Bai, Ting and Chen, Weijie and Yang, Cheng and Shi, Chuan},
  year = 2025,
  month = feb,
  number = {arXiv:2412.20036},
  eprint = {2412.20036},
  primaryclass = {cs},
  publisher = {arXiv},
  doi = {10.48550/arXiv.2412.20036},
  archiveprefix = {arXiv},
  langid = {american}
}

@inproceedings{AutoDebias,
  title = {{{AutoDebias}}: {{Learning}} to {{Debias}} for {{Recommendation}}},
  booktitle = {Proceedings of the 44th {{International ACM SIGIR Conference}} on {{Research}} and {{Development}} in {{Information Retrieval}}},
  author = {Chen, Jiawei and Dong, Hande and Qiu, Yang and He, Xiangnan and Xin, Xin and Chen, Liang and Lin, Guli and Yang, Keping},
  year = 2021,
  month = jul,
  pages = {21--30},
  publisher = {ACM},
  address = {Virtual Event Canada},
  doi = {10.1145/3404835.3462919},
  isbn = {978-1-4503-8037-9},
  langid = {english}
}

@article{chenBiasDebiasRecommender2023,
  title = {Bias and {{Debias}} in {{Recommender System}}: {{A Survey}} and {{Future Directions}}},
  author = {Chen, Jiawei and Dong, Hande and Wang, Xiang and Feng, Fuli and Wang, Meng and He, Xiangnan},
  year = 2023,
  month = jul,
  journal = {ACM Transactions on Information Systems},
  volume = {41},
  number = {3},
  pages = {1--39},
  issn = {1046-8188, 1558-2868},
  doi = {10.1145/3564284},
  langid = {english}
}

@inproceedings{InterD,
  title = {Interpolative {{Distillation}} for {{Unifying Biased}} and {{Debiased Recommendation}}},
  booktitle = {Proceedings of the 45th {{International ACM SIGIR Conference}} on {{Research}} and {{Development}} in {{Information Retrieval}}},
  author = {Ding, Sihao and Feng, Fuli and He, Xiangnan and Jin, Jinqiu and Wang, Wenjie and Liao, Yong and Zhang, Yongdong},
  year = 2022,
  month = jul,
  pages = {40--49},
  publisher = {ACM},
  address = {Madrid Spain},
  doi = {10.1145/3477495.3532002},
  isbn = {978-1-4503-8732-3},
  langid = {english}
}

@inproceedings{gaoKuaiRandUnbiasedSequential2022,
  title = {{{KuaiRand}}: {{An Unbiased Sequential Recommendation Dataset}} with {{Randomly Exposed Videos}}},
  booktitle = {Proceedings of the 31st {{ACM International Conference}} on {{Information}} \& {{Knowledge Management}}},
  author = {Gao, Chongming and Li, Shijun and Zhang, Yuan and Chen, Jiawei and Li, Biao and Lei, Wenqiang and Jiang, Peng and He, Xiangnan},
  year = 2022,
  month = oct,
  eprint = {2208.08696},
  primaryclass = {cs},
  pages = {3953--3957},
  doi = {10.1145/3511808.3557624},
  archiveprefix = {arXiv},
  langid = {american}
}

@inproceedings{guoEnhancedDoublyRobust2021,
  title = {Enhanced {{Doubly Robust Learning}} for {{Debiasing Post-Click Conversion Rate Estimation}}},
  booktitle = {Proceedings of the 44th {{International ACM SIGIR Conference}} on {{Research}} and {{Development}} in {{Information Retrieval}}},
  author = {Guo, Siyuan and Zou, Lixin and Liu, Yiding and Ye, Wenwen and Cheng, Suqi and Wang, Shuaiqiang and Chen, Hechang and Yin, Dawei and Chang, Yi},
  year = 2021,
  month = jul,
  pages = {275--284},
  publisher = {ACM},
  address = {Virtual Event Canada},
  doi = {10.1145/3404835.3462917},
  isbn = {978-1-4503-8037-9},
  langid = {english}
}

@misc{hintonDistillingKnowledgeNeural2015,
  title = {Distilling the {{Knowledge}} in a {{Neural Network}}},
  author = {Hinton, Geoffrey and Vinyals, Oriol and Dean, Jeff},
  year = 2015,
  month = mar,
  number = {arXiv:1503.02531},
  eprint = {1503.02531},
  primaryclass = {stat},
  publisher = {arXiv},
  doi = {10.48550/arXiv.1503.02531},
  archiveprefix = {arXiv}
}

@inproceedings{jiangDegenerateFeedbackLoops2019,
  title = {Degenerate {{Feedback Loops}} in {{Recommender Systems}}},
  booktitle = {Proceedings of the 2019 {{AAAI}}/{{ACM Conference}} on {{AI}}, {{Ethics}}, and {{Society}}},
  author = {Jiang, Ray and Chiappa, Silvia and Lattimore, Tor and Gy{\"o}rgy, Andr{\'a}s and Kohli, Pushmeet},
  year = 2019,
  month = jan,
  eprint = {1902.10730},
  primaryclass = {stat},
  pages = {383--390},
  doi = {10.1145/3306618.3314288},
  archiveprefix = {arXiv}
}

@inproceedings{krishnanAdversarialApproachImprove2018,
  title = {An {{Adversarial Approach}} to {{Improve Long-Tail Performance}} in {{Neural Collaborative Filtering}}},
  booktitle = {Proceedings of the 27th {{ACM International Conference}} on {{Information}} and {{Knowledge Management}}},
  author = {Krishnan, Adit and Sharma, Ashish and Sankar, Aravind and Sundaram, Hari},
  year = 2018,
  month = oct,
  pages = {1491--1494},
  publisher = {ACM},
  address = {Torino Italy},
  doi = {10.1145/3269206.3269264},
  isbn = {978-1-4503-6014-2},
  langid = {english}
}

@inproceedings{liangModelingUserExposure2016,
  title = {Modeling {{User Exposure}} in {{Recommendation}}},
  booktitle = {Proceedings of the 25th {{International Conference}} on {{World Wide Web}}},
  author = {Liang, Dawen and Charlin, Laurent and McInerney, James and Blei, David M.},
  year = 2016,
  month = apr,
  pages = {951--961},
  publisher = {International World Wide Web Conferences Steering Committee},
  address = {Montr\'eal Qu\'ebec Canada},
  doi = {10.1145/2872427.2883090},
  isbn = {978-1-4503-4143-1},
  langid = {english}
}

@inproceedings{liCausalRecommendationMachine2025,
  title = {Causal Recommendation via Machine Unlearning with a Few Unbiased Data},
  booktitle = {{{AAAI}} 2025 {{Workshop}} on {{Artificial Intelligence}} with {{Causal Techniques}}},
  author = {Li, Meng and Sui, Haochen},
  year = 2025,
  volume = {2},
  langid = {american}
}

@inproceedings{liRelaxingAccurateImputation2024,
  title = {Relaxing the Accurate Imputation Assumption in Doubly Robust Learning for Debiased Collaborative Filtering},
  booktitle = {Forty-First {{International Conference}} on {{Machine Learning}}},
  author = {Li, Haoxuan and Zheng, Chunyuan and Wang, Shuyi and Wu, Kunhan and Wang, Eric and Wu, Peng and Geng, Zhi and Chen, Xu and Zhou, Xiao-Hua},
  year = 2024
}

@misc{liTDRCLTargetedDoubly2023,
  title = {{{TDR-CL}}: {{Targeted Doubly Robust Collaborative Learning}} for {{Debiased Recommendations}}},
  author = {Li, Haoxuan and Lyu, Yan and Zheng, Chunyuan and Wu, Peng},
  year = 2023,
  month = mar,
  number = {arXiv:2203.10258},
  eprint = {2203.10258},
  primaryclass = {cs},
  publisher = {arXiv},
  doi = {10.48550/arXiv.2203.10258},
  archiveprefix = {arXiv}
}

@inproceedings{liuGeneralKnowledgeDistillation2020,
  title = {A {{General Knowledge Distillation Framework}} for {{Counterfactual Recommendation}} via {{Uniform Data}}},
  booktitle = {Proceedings of the 43rd {{International ACM SIGIR Conference}} on {{Research}} and {{Development}} in {{Information Retrieval}}},
  author = {Liu, Dugang and Cheng, Pengxiang and Dong, Zhenhua and He, Xiuqiang and Pan, Weike and Ming, Zhong},
  year = 2020,
  month = jul,
  pages = {831--840},
  publisher = {ACM},
  address = {Virtual Event China},
  doi = {10.1145/3397271.3401083},
  isbn = {978-1-4503-8016-4},
  langid = {english}
}

@inproceedings{liuRatingDistributionCalibration2022,
  title = {Rating {{Distribution Calibration}} for {{Selection Bias Mitigation}} in {{Recommendations}}},
  booktitle = {Proceedings of the {{ACM Web Conference}} 2022},
  author = {Liu, Haochen and Tang, Da and Yang, Ji and Zhao, Xiangyu and Liu, Hui and Tang, Jiliang and Cheng, Youlong},
  year = 2022,
  month = apr,
  series = {{{WWW}} '22},
  pages = {2048--2057},
  publisher = {Association for Computing Machinery},
  address = {New York, NY, USA},
  doi = {10.1145/3485447.3512078},
  isbn = {978-1-4503-9096-5}
}

@misc{mansouryFeedbackLoopBias2020,
  title = {Feedback {{Loop}} and {{Bias Amplification}} in {{Recommender Systems}}},
  author = {Mansoury, Masoud and Abdollahpouri, Himan and Pechenizkiy, Mykola and Mobasher, Bamshad and Burke, Robin},
  year = 2020,
  month = jul,
  number = {arXiv:2007.13019},
  eprint = {2007.13019},
  primaryclass = {cs},
  publisher = {arXiv},
  doi = {10.48550/arXiv.2007.13019},
  archiveprefix = {arXiv}
}

@misc{marlinCollaborativeFilteringMissing2012,
  title = {Collaborative {{Filtering}} and the {{Missing}} at {{Random Assumption}}},
  author = {Marlin, Benjamin and Zemel, Richard S. and Roweis, Sam and Slaney, Malcolm},
  year = 2012,
  month = jun,
  number = {arXiv:1206.5267},
  eprint = {1206.5267},
  primaryclass = {cs},
  publisher = {arXiv},
  doi = {10.48550/arXiv.1206.5267},
  archiveprefix = {arXiv}
}

@article{rowlandFilterBubbleWhat2011,
  title = {The Filter Bubble: {{What}} the Internet Is Hiding from You},
  author = {Rowland, Fred},
  year = 2011,
  journal = {portal: Libraries and the Academy},
  volume = {11},
  number = {4},
  pages = {1009--1011},
  publisher = {Johns Hopkins University Press}
}

@misc{saitoUnbiasedRecommenderLearning2020,
  title = {Unbiased {{Recommender Learning}} from {{Missing-Not-At-Random Implicit Feedback}}},
  author = {Saito, Yuta and Yaginuma, Suguru and Nishino, Yuta and Sakata, Hayato and Nakata, Kazuhide},
  year = 2020,
  month = feb,
  number = {arXiv:1909.03601},
  eprint = {1909.03601},
  primaryclass = {stat},
  publisher = {arXiv},
  doi = {10.48550/arXiv.1909.03601},
  archiveprefix = {arXiv}
}

@inproceedings{IPS,
  title = {Recommendations as Treatments: {{Debiasing}} Learning and Evaluation},
  booktitle = {International Conference on Machine Learning},
  author = {Schnabel, Tobias and Swaminathan, Adith and Singh, Ashudeep and Chandak, Navin and Joachims, Thorsten},
  year = 2016,
  pages = {1670--1679},
  publisher = {PMLR}
}

@inproceedings{wangCombatingSelectionBiases2021,
  title = {Combating {{Selection Biases}} in {{Recommender Systems}} with a {{Few Unbiased Ratings}}},
  booktitle = {Proceedings of the 14th {{ACM International Conference}} on {{Web Search}} and {{Data Mining}}},
  author = {Wang, Xiaojie and Zhang, Rui and Sun, Yu and Qi, Jianzhong},
  year = 2021,
  month = mar,
  series = {{{WSDM}} '21},
  pages = {427--435},
  publisher = {Association for Computing Machinery},
  address = {New York, NY, USA},
  doi = {10.1145/3437963.3441799},
  isbn = {978-1-4503-8297-7}
}

@inproceedings{wangDoublyRobustJoint2019,
  title = {Doubly Robust Joint Learning for Recommendation on Data Missing Not at Random},
  booktitle = {International {{Conference}} on {{Machine Learning}}},
  author = {Wang, Xiaojie and Zhang, Rui and Sun, Yu and Qi, Jianzhong},
  year = 2019,
  pages = {6638--6647},
  publisher = {PMLR}
}

@inproceedings{wangInvariantPreferenceLearning2022,
  title = {Invariant {{Preference Learning}} for {{General Debiasing}} in {{Recommendation}}},
  booktitle = {Proceedings of the 28th {{ACM SIGKDD Conference}} on {{Knowledge Discovery}} and {{Data Mining}}},
  author = {Wang, Zimu and He, Yue and Liu, Jiashuo and Zou, Wenchao and Yu, Philip S. and Cui, Peng},
  year = 2022,
  month = aug,
  series = {{{KDD}} '22},
  pages = {1969--1978},
  publisher = {Association for Computing Machinery},
  address = {New York, NY, USA},
  doi = {10.1145/3534678.3539439},
  isbn = {978-1-4503-9385-0},
  langid = {american}
}

@misc{zhangDebiasingRecommendationLearning2023a,
  title = {Debiasing {{Recommendation}} by {{Learning Identifiable Latent Confounders}}},
  author = {Zhang, Qing and Zhang, Xiaoying and Liu, Yang and Wang, Hongning and Gao, Min and Zhang, Jiheng and Guo, Ruocheng},
  year = 2023,
  month = feb,
  journal = {arXiv.org},
  howpublished = {https://arxiv.org/abs/2302.05052v2},
  langid = {english}
}

@misc{zhangGeneralDebiasingGraphbased2024a,
  title = {General {{Debiasing}} for {{Graph-based Collaborative Filtering}} via {{Adversarial Graph Dropout}}},
  author = {Zhang, An and Ma, Wenchang and Wei, Pengbo and Sheng, Leheng and Wang, Xiang},
  year = 2024,
  month = feb,
  journal = {arXiv.org},
  howpublished = {https://arxiv.org/abs/2402.13769v1},
  langid = {english}
}

@misc{collinsStudyPositionBias2018,
  title = {A {{Study}} of {{Position Bias}} in {{Digital Library Recommender Systems}}},
  author = {Collins, Andrew and Tkaczyk, Dominika and Aizawa, Akiko and Beel, Joeran},
  year = 2018,
  month = feb,
  journal = {arXiv.org},
  howpublished = {https://arxiv.org/abs/1802.06565v1},
  langid = {english}
}

@inproceedings{zieglerImprovingRecommendationLists2005,
  title = {Improving Recommendation Lists through Topic Diversification},
  booktitle = {Proceedings of the 14th International Conference on {{World Wide Web}}  - {{WWW}} '05},
  author = {Ziegler, Cai-Nicolas and McNee, Sean M. and Konstan, Joseph A. and Lausen, Georg},
  year = 2005,
  pages = {22},
  publisher = {ACM Press},
  address = {Chiba, Japan},
  doi = {10.1145/1060745.1060754},
  copyright = {https://www.acm.org/publications/policies/copyright\_policy\#Background},
  isbn = {978-1-59593-046-0},
  langid = {english}
}

@article{kaminskasDiversitySerendipityNovelty2017,
  title = {Diversity, {{Serendipity}}, {{Novelty}}, and {{Coverage}}: {{A Survey}} and {{Empirical Analysis}} of {{Beyond-Accuracy Objectives}} in {{Recommender Systems}}},
  author = {Kaminskas, Marius and Bridge, Derek},
  year = 2017,
  month = mar,
  journal = {ACM Transactions on Interactive Intelligent Systems},
  volume = {7},
  number = {1},
  pages = {1--42},
  issn = {2160-6455, 2160-6463},
  doi = {10.1145/2926720},
  langid = {english}
}

@inproceedings{steckEvaluationRecommendationsRatingprediction2013,
  title = {Evaluation of Recommendations: Rating-Prediction and Ranking},
  booktitle = {Proceedings of the 7th {{ACM}} Conference on {{Recommender}} Systems},
  author = {Steck, Harald},
  year = 2013,
  month = oct,
  pages = {213--220},
  publisher = {ACM},
  address = {Hong Kong China},
  doi = {10.1145/2507157.2507160},
  isbn = {978-1-4503-2409-0},
  langid = {english}
}

@inproceedings{yiSamplingbiascorrectedNeuralModeling2019,
  title = {Sampling-Bias-Corrected Neural Modeling for Large Corpus Item Recommendations},
  booktitle = {Proceedings of the 13th {{ACM Conference}} on {{Recommender Systems}}},
  author = {Yi, Xinyang and Yang, Ji and Hong, Lichan and Cheng, Derek Zhiyuan and Heldt, Lukasz and Kumthekar, Aditee and Zhao, Zhe and Wei, Li and Chi, Ed},
  year = 2019,
  month = sep,
  series = {{{RecSys}} '19},
  pages = {269--277},
  publisher = {Association for Computing Machinery},
  address = {New York, NY, USA},
  doi = {10.1145/3298689.3346996},
  isbn = {978-1-4503-6243-6}
}

@inproceedings{marlinCollaborativePredictionRanking2009,
  title = {Collaborative Prediction and Ranking with Non-Random Missing Data},
  booktitle = {Proceedings of the Third {{ACM}} Conference on {{Recommender}} Systems},
  author = {Marlin, Benjamin M. and Zemel, Richard S.},
  year = 2009,
  month = oct,
  pages = {5--12},
  publisher = {ACM},
  address = {New York New York USA},
  doi = {10.1145/1639714.1639717},
  isbn = {978-1-60558-435-5},
  langid = {english}
}

@article{korenMatrixFactorizationTechniques2009,
  title = {Matrix {{Factorization Techniques}} for {{Recommender Systems}}},
  author = {Koren, Yehuda and Bell, Robert and Volinsky, Chris},
  year = 2009,
  month = aug,
  journal = {Computer},
  volume = {42},
  number = {8},
  pages = {30--37},
  issn = {1558-0814},
  doi = {10.1109/MC.2009.263}
}

@article{zhuAdversarialPropensityWeighting,
  title = {Adversarial {{Propensity Weighting}} for {{Debiasing}} in {{Collaborative Filtering}}},
  author = {Zhu, Kuiyu and Qin, Tao and Wang, Pinghui and Wang, Xin},
  langid = {english}
}

@inproceedings{weiModelAgnosticCounterfactualReasoning2021,
  title = {Model-{{Agnostic Counterfactual Reasoning}} for {{Eliminating Popularity Bias}} in {{Recommender System}}},
  booktitle = {Proceedings of the 27th {{ACM SIGKDD Conference}} on {{Knowledge Discovery}} \& {{Data Mining}}},
  author = {Wei, Tianxin and Feng, Fuli and Chen, Jiawei and Wu, Ziwei and Yi, Jinfeng and He, Xiangnan},
  year = 2021,
  month = aug,
  pages = {1791--1800},
  publisher = {ACM},
  address = {Virtual Event Singapore},
  doi = {10.1145/3447548.3467289},
  isbn = {978-1-4503-8332-5},
  langid = {english}
}

\end{document}